%% file: ANA-HDBS-2018-17-PAPER.tex
\documentclass[cernpreprint,USenglish,atlasdraft=true,texlive=2017,texmf]{atlasdoc}
\usepackage{todonotes}
\DeclareBoolOption[true]{LANGEDIT}
\LEcontact{Chip Brock brockr@msu.edu}

\usepackage[subfigure=true]{atlaspackage}
\usepackage{afterpage,lipsum}
\usepackage{atlasbiblatex}
\usepackage{atlasphysics}
 
\graphicspath{{logos/}{figures/}}

\PreprintIdNumber{CERN-EP-2020-128}
\AtlasTitle{Search for Heavy Resonances Decaying into a Photon and a Hadronically Decaying Higgs Boson in $pp$ Collisions at
$\sqrt{s}=13\,\tev$ with the ATLAS Detector}

\AtlasJournalRef{Phys. Rev. Lett}
\AtlasDOI{10.1103/PhysRevLett.125.251802}
 
\AtlasAbstract{
This Letter presents a search for the production of new heavy resonances decaying into a Higgs boson and a photon using
proton-proton collision data at  $\sqrt{s}=13$\,TeV collected by the ATLAS detector at the LHC.
The data correspond to an integrated luminosity of $139\,\mbox{fb}^{-1}$.  The analysis is performed by reconstructing
hadronically decaying Higgs boson $(H\to b\bar{b})$ candidates as single large-radius jets. A novel algorithm using information about the jet constituents
in the center-of-mass frame of the jet is implemented to identify the two $b$-quarks in the single jet.  No significant excess of events is
observed above the expected background.
Upper limits are set on the production cross-section times branching fraction for narrow spin-1 resonances decaying into a Higgs boson and a photon
in the resonance mass range from $0.7$ to $4$\,TeV, cross-sections times branching fractions are excluded
between $11.6\,\mbox{fb}$ and $0.11\,\mbox{fb}$ at a 95\% confidence level.
}
 
\author{ATLAS Collaboration}
 
\AtlasRefCode{HDBS-2018-17}

\hypersetup{pdftitle={ATLAS document},pdfauthor={The ATLAS Collaboration}}
 
\begin{document}
 
\maketitle

Many extensions to the standard model, such as technicolor~\cite{Eichten:2007sx}, little Higgs~\cite{ArkaniHamed:2001nc}, or 
a more complex Higgs sector~\cite{Low:2011gn},  predict new massive bosons. Some of these bosons may decay into a Higgs boson and a
photon at the one-loop level~\cite{Dobrescu:2017sue}. Searches for such particles have been carried out by both the ATLAS~\cite{EXOT-2016-30} and
CMS~\cite{CMS-EXO-17-019} Collaborations at the Large Hadron Collider (LHC).
 
This Letter reports on a generic search for a narrow, neutral, spin-1 boson ($Z'$) that decays into a photon and a Higgs boson.
The Higgs boson subsequently decays hadronically
as $H\to b\bar{b}$, where the hadronic products from both $b$ quarks are reconstructed as a single large-radius jet.
The analysis uses data from $\sqrt{s}=13\,\tev$ proton--proton ($pp$) collisions that were recorded by the ATLAS detector
from 2015 to 2018 with a single-photon trigger~\cite{TRIG-2018-05}, corresponding to an integrated luminosity of  $139\,\ifb$.
The single-photon trigger uses loose photon identification requirements based on calorimetric shower-shape variables~\cite{PERF-2013-04}
and imposes a transverse momentum threshold of $140\,\gev$. It is fully efficient for events passing the offline analysis selection.
The search identifies the two $b$ quarks in the single jet by using a novel methodology based on information about the jet
constituents calculated in the center-of-mass frame of the jet. This technique significantly improves the search sensitivity
compared to the previous ATLAS~\cite{EXOT-2016-30} and CMS~\cite{CMS-EXO-17-019} analyses, in addition to the gains
from the larger data sample.

The ATLAS detector~\cite{PERF-2007-01,,PIX-2018-001} is a general-purpose particle detector with a cylindrical geometry\footnote{ATLAS uses a right-handed
coordinate system with its origin at the nominal interaction point in the center of the detector and the $z$-axis along the beam pipe. The
$x$-axis points to the center of the LHC ring, and the $y$-axis points upward. Cylindrical coordinates $(r, \phi)$ are used in the transverse plane,
$\phi$ being the azimuthal angle around the $z$--axis. The pseudorapidity is defined in terms of the polar angle $\theta$ as $\eta = - \ln \tan (\theta/2)$.
The distance between two objects in $\eta$--$\phi$ space is $\Delta R = \sqrt{(\Delta \eta)^2 + (\Delta \phi)^2}$.
Transverse momentum is defined by $\pt = p \sin \theta$.
}.
It consists of an inner detector
surrounded by a superconducting solenoid that produces a 2\,T magnetic field,
electromagnetic and hadronic calorimeters, and a muon spectrometer with a toroidal magnetic field. The inner detector provides precision tracking
of charged particles with pseudorapidity $|\eta| < 2.5$. The calorimeter system covers the pseudorapidity range $|\eta| < 4.9$.
It comprises sampling calorimeters with either liquid argon or scintillator tiles as the active medium.
A two-level trigger system
accepts events from the 40\,MHz bunch crossings at a rate of 1\,kHz for off-line analysis.
 
Monte Carlo (MC) simulated events are used to optimize the event selection and to help validate the analysis.
The signal samples, with decays of $Z' \rightarrow H\gamma$ at the one-loop level~\cite{Dobrescu:2017sue}, were generated for eight different mass points
in a range from $700$ to $4000\,\gev$ via quark-antiquark annihilation, $q\bar{q}\to Z'\to H\gamma$, using the \MADGRAPH leading-order (LO) v2.6.2
generator~\cite{Alwall:2014hca} interfaced to
\PYTHIAV{8.235}~\cite{Sjostrand:2007gs} with the NNPDF23LO parton distribution functions (PDFs)~\cite{Ball:2012cx} for both generators and the
A14 set of tuned parameters~\cite{ATL-PHYS-PUB-2014-021} for the underlying event. The total decay widths of the $Z'$ resonances were set to $4.2\,\mev$,
which is much smaller than the experimental mass resolution, which varies from around $35\,\gev$ at the $700\,\gev$ signal mass point to $150\,\gev$ at the $4000\,\gev$ signal mass point.
The dominant SM background arises from events with prompt photons produced in association with jets ($\gamma$+jets). Less dominant
SM backgrounds include a prompt photon produced in association with a $W/Z$ boson ($W/Z$+$\gamma$) or a top--anti-top quark pair ($t\bar{t}$+$\gamma$).
The MC sample of $\gamma$+jets events was simulated using the \SHERPAV{2.2.2} generator~\cite{Bothmann:2019yzt}
with up to two additional parton emissions at next-to-leading-order (NLO) accuracy and up to four additional partons at LO accuracy using Comix~\cite{Gleisberg:2008fv}
and OpenLoops~\cite{Cascioli:2011va}. The events were then merged with the \SHERPA parton
shower~\cite{Schumann:2007mg} using the ME+PS@NLO prescription~\cite{Hoeche:2009rj}. Samples are generated using the
NNPDF3.0nnlo PDF set~\cite{Ball:2014uwa}, along with the dedicated set of
tuned parton-shower parameters developed by the \SHERPA authors.
The $W/Z$+$\gamma$ events were modeled with \SHERPAV{2.1.1} at LO with the CT10 PDFs~\cite{Lai:2010vv} for both generators and the underlying event.
The $t\bar{t}$+$\gamma$ events were simulated using \MGMCatNLOV{v2.2.3} at LO with the CTEQ6L1 PDFs~\cite{Pumplin:2002vw}, then interfaced to \PYTHIAV{8.186}  with
the A14 parameter tune and the NNPDF23LO PDFs. In the signal samples and $t\bar{t}$+$\gamma$ background sample, \textsc{EvtGen}~\cite{Lange:2001uf} was used to model charm
and $b$-hadron decays.
The effect of multiple $pp$ interactions in the same and neighboring bunch crossings (pileup) is included by overlaying minimum-bias events simulated with
\PYTHIAV{8.186} on each event of interest in all samples. The generated samples were processed through a $\GEANT4$-based detector
simulation~\cite{Agostinelli:2002hh,SOFT-2010-01} and the same ATLAS reconstruction software as the data.
 
An event is selected if it contains a $H\to b\bar{b}$ candidate and at least one isolated  photon that satisfies the ``tight'' identification criteria~\cite{ATL-PHYS-PUB-2016-014,ATL-PHYS-PUB-2016-015}.
A selected photon must have transverse momentum (\pt) greater than $200\,\gev$ and be within the calorimeter barrel region\,$|\eta|<1.37$.
Each $H\to b\bar{b}$ candidate is reconstructed as a single jet using the anti-$k_t$ algorithm~\cite{Cacciari:2008gp, Cacciari:2011ma} with a large radius parameter ($R=1.0$), hereafter referred to as a large-$R$ jet ($J$).  The large-$R$ jets are formed from topological energy clusters (topoclusters)~\cite{PERF-2014-07} in the
calorimeter and are trimmed~\cite{Krohn:2009th} to mitigate the effects of pileup and soft radiation. The large-$R$ jet constituents are reclustered into subjets using the
$k_t$ algorithm~\cite{Catani:1993hr} with $R=0.2$, and the subjets that carry less than 5\% of the \pt\ of the original large-$R$ jet are removed.
To overcome the limited angular resolution of the calorimeter, the mass of a large-$R$ jet ($m_J$) is computed using a combination
of calorimeter and tracking information~\cite{ATLAS-CONF-2016-035}. Large-$R$ jets are required to have $\pt>200\,\gev$, $|\eta|<2.0$,
$50\,\gev<m_J<200\,\gev$,  and an angular separation of $\Delta R > 1.0$ from photon candidates. For the baseline event selection, at least one large-$R$ jet and one photon are required
to pass the selection described above.
 
The photon and large-$R$ jet with the highest \pt\  in an event are combined to form a resonance candidate.
The invariant mass of the resonance candidate ($m_{J\gamma}$) is used to distinguish signal from background.
In addition, the large-$R$ jet mass must be consistent with the Higgs boson mass ($m_H = 125.80\,\gev$), 
$m_H-\Delta_{m,D}<m_J<m_H+\Delta_{m,U}$. The parameters $\Delta_{m,D}$
and $\Delta_{m,U}$ are determined by maximizing the
search sensitivity $\epsilon/(\sqrt{B}+3/2)$~\cite{Punzi:2003bu}, where $\epsilon$ is the resonance signal selection efficiency  and
$B$ is the number of background events, as estimated from MC samples, within the resonance mass window, $|m_{J\gamma}-\bar{m}_{Z'}|<2\sigma_{m_{Z'}}$. 
Here $\bar{m}_{Z'}$ and $\sigma_{m_{Z'}}$ are the peak position and the core resolution of the
reconstructed $m_{J\gamma}$ distribution of the $Z'\to H\gamma$ signal MC events, respectively.
The above procedure is performed separately for each $m_{Z'}$ hypothesis. The optimized parameters $\Delta_{m,D}$
and $\Delta_{m,U}$ are then parameterized by fourth-order polynomial functions of the large-$R$ jet \pt.
The optimized mass window of the large-$R$ jets varies from around $[100, 130]\,\gev$ at $\pt=0.5\,\tev$ to $[90, 160]\,\gev$ at
$\pt=2\,\tev$.
 
To further reduce the background, a novel algorithm~\cite{Chen:2015tma,Chen:2011ah,STDM-2012-21} is applied to the large-$R$ jet to identify the
two $b$ quarks that originated from the Higgs boson. It uses the kinematics of the jet constituents in the  center-of-mass (c.m.) frame of the
large-$R$ jet (jet rest frame), where the final products of a two-body $H\to b\bar{b}$ decay can be easily separated into a back-to-back topology.
In this approach, the topoclusters of  the large-$R$  jet and the tracks associated with the jet are boosted to
the large-$R$ jet's rest frame. In the jet rest frame, the topoclusters of the large-$R$  jet are reclustered using the EEkT jet algorithm~\cite{catani1991new} to
form exactly two c.m. subjets, assumed to originate from the Higgs boson decay. A track is considered to be associated with a c.m. subjet if the
opening angle $\Delta\theta$ between the track and the c.m. subjet, calculated in the jet rest frame, satisfies the requirement that $2\times (1-\cos\Delta\theta)<0.8$.
The c.m. subjets and their associated
tracks are then boosted back to the laboratory frame and the standard ATLAS $b$-tagging algorithm based on a multivariate technique,
MV2c10~\cite{PERF-2012-04,ATL-PHYS-PUB-2016-012}, is applied to each c.m. subjet to identify those containing a $b$ hadron (called c.m. $b$ subjets).
For this analysis, the working point of the MV2c10 tagger output is chosen to have an overall efficiency of 77\%.
This was determined using simulated Randall-Sundrum graviton~\cite{Randall:1999ee} ($G\to HH$, $H\to b\bar{b}$) events,
in which the \pt\ distribution of the large-$R$ jets that contain a Higgs boson is reweighted to match the inclusive jet \pt\ distribution observed
in data~\cite{ATL-PHYS-PUB-2017-010}.
Compared to the previous method used to identify $H\to b\bar{b}$ reconstructed as large-$R$ jets, MC studies~\cite{ATL-PHYS-PUB-2017-010}
show that $b$-tagging based on c.m. subjets can reject more background than the $b$ tagging based on the other subjet algorithm at a given
signal identification efficiency: by 20\%--50\% for large-$R$ jets with $\pt\le 1.5\,\tev$ and up to a factor of 10 or more for large-$R$ jets with $\pt> 1.5\,\tev$.
Among several tagging techniques~\cite{ATL-PHYS-PUB-2017-010}  developed to improve the identification of $H\to b\bar{b}$ with $\pt> 1\,\tev$,
the c.m. algorithm typically rejects 20\% more background at a given signal efficiency.
 
Studies using MC simulated events show that the correlation between the $b$-tagging efficiencies of two c.m. $b$ subjets is negligible,
and thus the $b$-tagging efficiency of each c.m. $b$ subjet in a large-$R$ jet can be calibrated using  boosted hadronic top-quark decays
$t\to Wb$ from $t\bar{t}\to WbW\bar{b}$ events where one $W$ boson decays hadronically and the other decays leptonically.
The hadronic products of the boosted $t\to Wb$ decay are reconstructed as a single large-$R$ jet, in which exactly two c.m. subjets are
reconstructed in the jet rest frame: one corresponding to the $b$ quark, and the other corresponding to the $W$ boson.
MC studies show that the $b$-tagging performance is almost identical for c.m. $b$ subjets in the boosted hadronic top-quark decay events and $H\to b\bar{b}$ events.
A standard combinatorial likelihood approach~\cite{ATLAS-CONF-2014-004} is applied to extract the c.m. $b$-subjet tagging efficiency
in order to calculate an MC-to-data scale factor, defined as the ratio of the c.m. $b$-subjet tagging efficiencies measured in data and simulated
$t\bar{t}$ events~\cite{ATL-PHYS-PUB-2016-004}.  The scale factor is found to be consistent with unity within its uncertainty
and has no significant dependence on the kinematics of the c.m. subjet and the large-$R$ jet. The uncertainty
of the scale factor is about 5\%, dominated by the systematic uncertainties such as the dependence of
the calibration scale factor on the choice of the $t\bar{t}$ MC generators, and the dependence of the
MV2c10~\cite{PERF-2012-04,ATL-PHYS-PUB-2016-012} $b$-tagging scale factors on the jet flavor.
 
The selected resonance candidates are retained for further analysis if one or both of the c.m. subjets in the large-$R$ jet pass the $b$-tagging requirement,
and are assigned to the single- or double-$b$-tagged category, respectively. Afterwards, optimizations of the selection requirements on the photon \pt\
($\pt^\gamma$) and the large-$R$ jet \pt\ ($\pt^J$) are carried out in sequence in order to further improve the search sensitivity.
The optimizations are performed separately for the selected events in the single- and double-$b$-tagged categories with the
same procedure as used for the large-$R$ jet mass-window optimization described above.  It yields $\pt^{\gamma}>\pt^0+a\times m_{J\gamma}$
and $\pt^J>0.8\times(\pt^0+a\times m_{J\gamma})$, where $\pt^0=12.0\,(121.8)\,\gev$ and $a=0.35$\,($0.22$) for  the selected events with
$m_{J\gamma}\leq2000\,(1500)\,\gev$ in the single-$b$-tagged (double-$b$-tagged) category. For events with $m_{J\gamma}>2000\,(1500)\,\gev$,
the selection requirements on the photon and the large-$R$ jet \pt\ are the same as those for events with $m_{J\gamma}=2000\,(1500)\,\gev$.
Depending on the resonance mass, the final signal efficiency in the single- and double-$b$-tagged categories varies between 10\% and 20\%.
 
The final discrimination between signal and background is achieved by a simultaneous fit to the $m_{J\gamma}$ distributions of the
selected data events in the single- and double-$b$-tagged categories. The signal probability density function (SPDF) is modeled as a sum of
a Crystal Ball function~\cite{CrystalBall} and a small Gaussian component that describes the tails produced by poorly reconstructed resonance candidates.
The SPDF parameters extracted from MC simulated events are interpolated as polynomial functions of the resonance mass up
to the third order.  Afterwards, the parameters of the SPDF at a given resonance mass are fixed to the values determined
using the parameterization.
The background probability density function (BPDF) is modeled as $B(m_{J\gamma})=(1-x)^{p_1}x^{p_2+p_3\log(x)}$~\cite{Harris:2011bh}, where $x=m_{J\gamma}/\sqrt{s}$, $\sqrt{s}=13\,\tev$ is the
center-of-mass energy, and the
three dimensionless shape parameters $p_1$, $p_2$, and $p_3$ are allowed to float in the fit. The choice of the BPDF is motivated and validated
by using control data samples containing events that satisfy all the signal selection criteria in either the single- or double-$b$-tagged
category, except for the $b$-tagging and large-$R$ jet mass requirements. The selected large-$R$ jet candidates in the control data samples
are required to have masses lying in sidebands, whose width varies from $10\,\gev$ to $30\,\gev$, separated from the Higgs boson signal band by $5\,\gev$,
and to have both of the c.m. subjets failing the $b$-tagging requirement at the 85\%-efficiency working point.
MC simulated events show that the background $m_{J\gamma}$ distributions
in the single- and double-$b$-tagged categories are well described by the events in the corresponding control sample.
 
The effect of systematic uncertainties from various sources was studied.
The uncertainty of the integrated luminosity is 1.7\,\%~\cite{DAPR-2013-01,Avoni:2018iuv}.
Uncertainties resulting from detector effects only affect the calculation of the signal selection efficiencies
since the background is estimated from the data. Those uncertainties include effects from
the energy and mass scales (2\%--6.5\%) of the large-$R$ jets~\cite{ATLAS-CONF-2017-063},
the large-$R$ jet energy resolution ($<0.2\%$) and mass resolution (18\%--30\%),
the trigger efficiencies ($<0.1\%$), the photon energy scale and resolution ($<2\%$)~\cite{ATL-PHYS-PUB-2016-015},
the photon reconstruction, identification and isolation efficiencies ($<0.1\%$)~\cite{PERF-2013-04},
the $b$-tagging efficiency of the c.m. subjet (3\%--15\%), and the pileup modeling ($<0.5\%$)~\cite{STDM-2015-05}.
In principle, the detector modeling may also affect the SPDF. However, such effects are found to be negligible.
The signal efficiency and acceptance are also affected by theoretical uncertainties, such as the PDF choice and
initial- and final-state radiation modeling. These are also found to be small ($< 5\%$ from the PDF, $<1\%$ from parton showering and $<1\%$ from renormalization-factorization scale).
The above systematic uncertainties degrade the final limits by 10\% at $700\,\gev$, increasing to around 20\% at $2.5\,\tev$ and back to 10\% at $4\,\tev$.
Another kind of uncertainty, referred to as the spurious signal, arises from a potential bias in the estimated
number of signal events due to the choice of background parameterization. It
was estimated by fitting the signal-plus-background model to control data sample $m_{J\gamma}$
distributions with a control region to signal region background-shape correction factor derived from simulation.
The absolute number of fitted signal events at a given $m_{Z'}$
hypothesis value is taken as the number of spurious-signal events, which varies from a few events in the low mass region to less than 0.1 events
in the high mass region, and is parameterized as an exponential function of $m_{Z'}$.
\afterpage{
\begin{figure*}[t]
\begin{center}
{\includegraphics[width=0.325\textwidth]{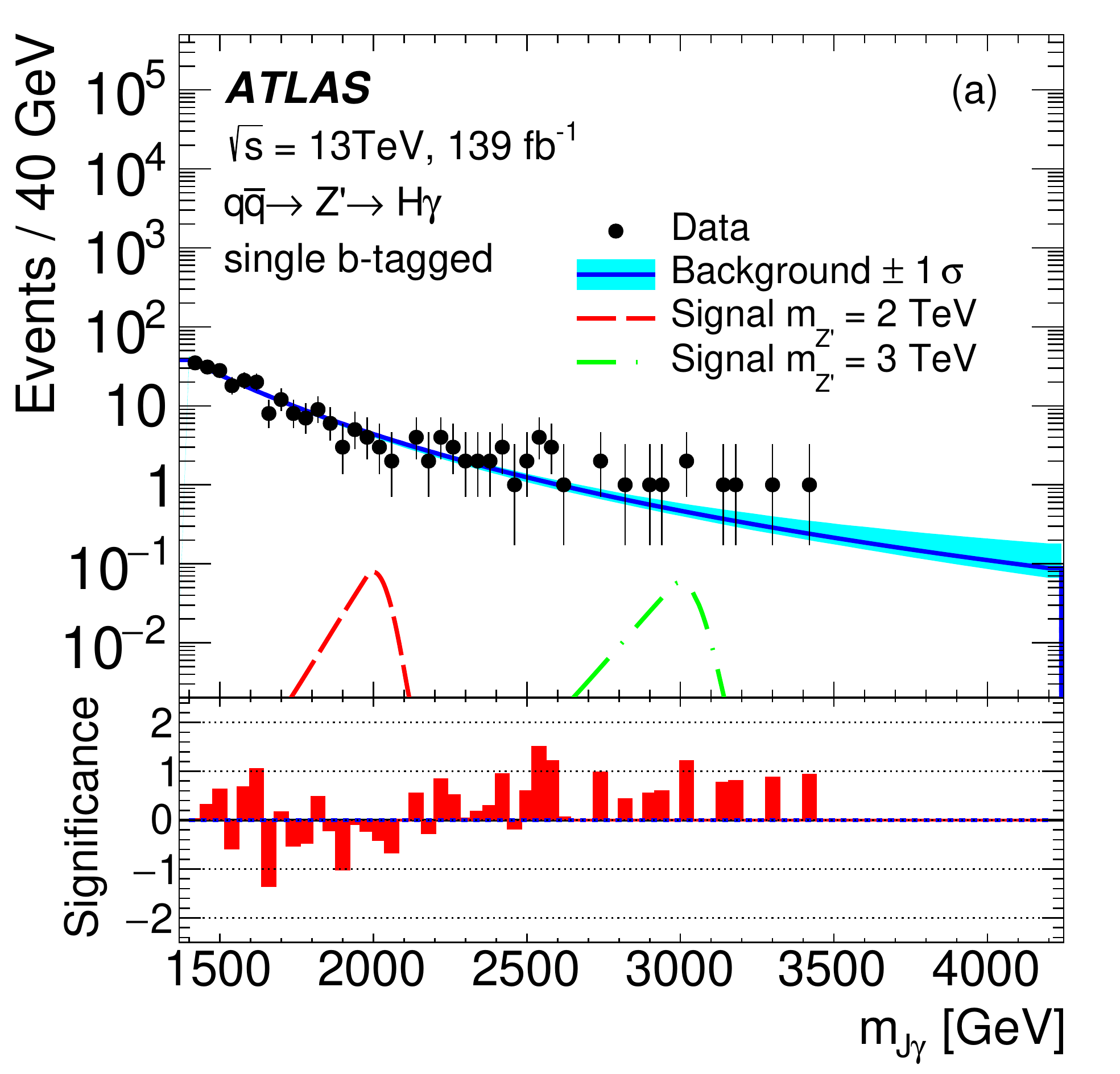}}
{\includegraphics[width=0.325\textwidth]{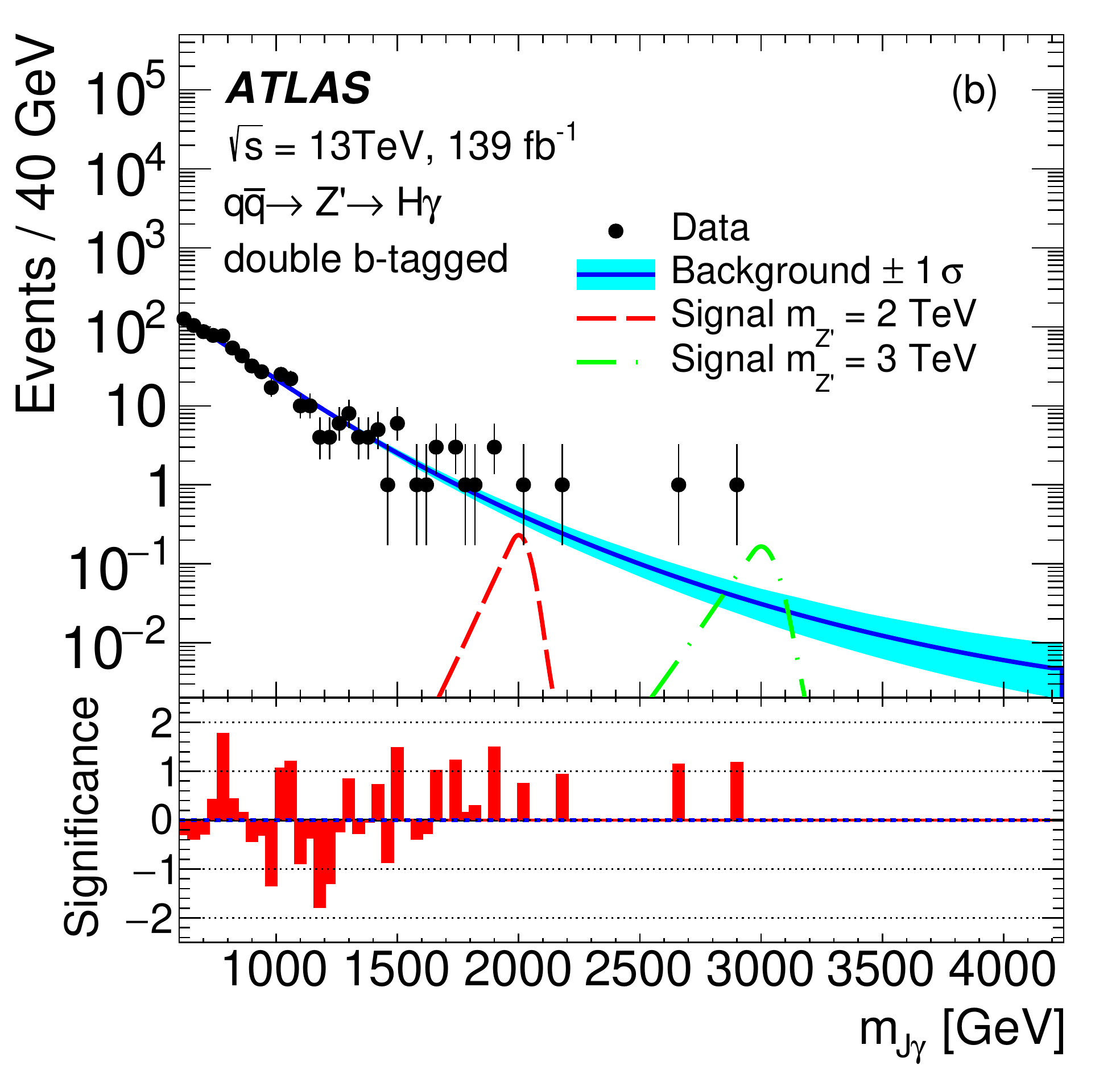}}
{\includegraphics[width=0.325\textwidth]{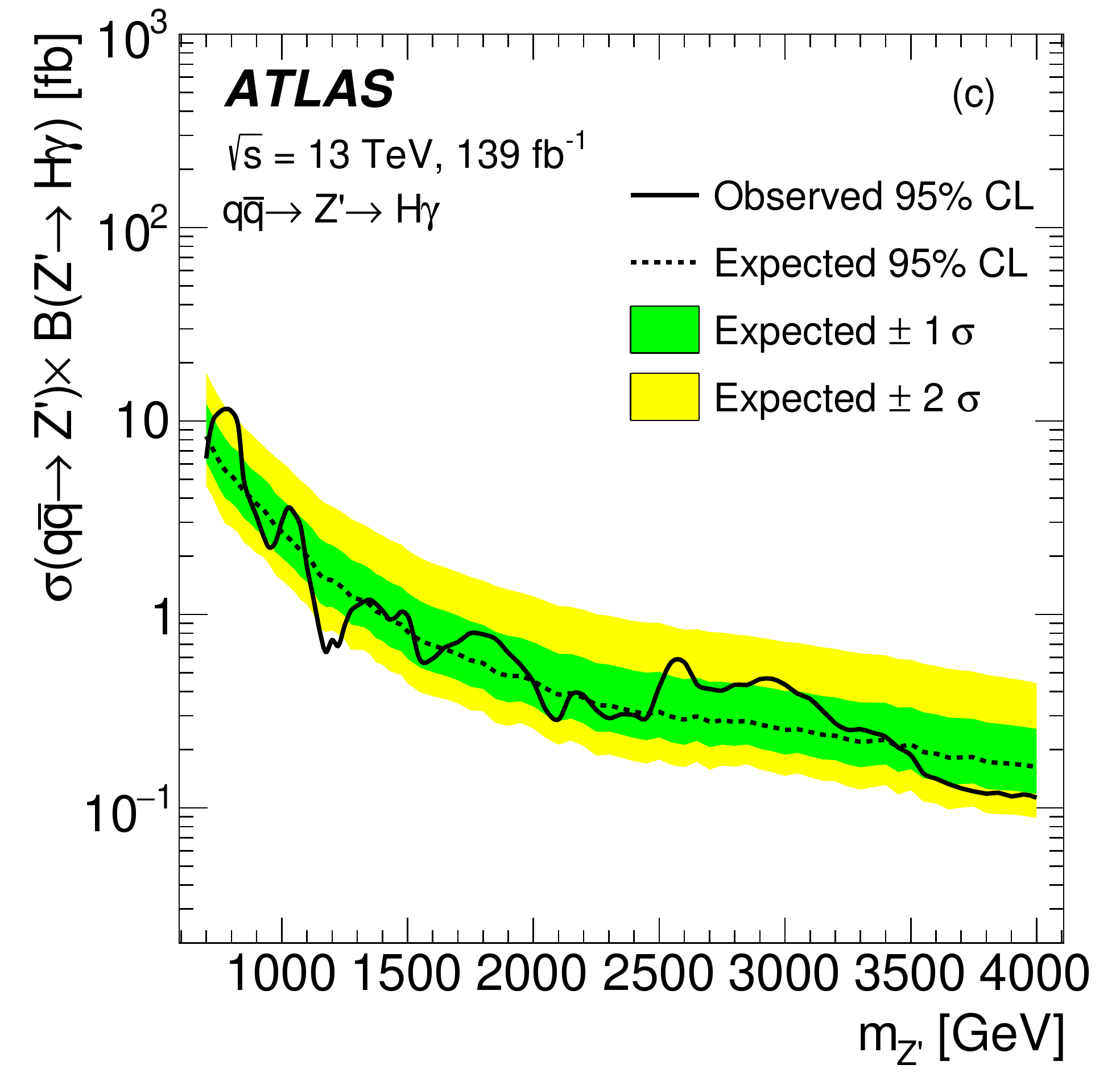}}
\caption{(a), (b)  Distribution of the reconstructed $m_{J\gamma}$ in the single- and double-$b$-tagged categories,
with the background-only fits shown by the solid lines.
The corresponding chi-square probabilities of the fits are 46\% and 23\%,
after rebinning requiring at least five events in each bin.
Hypothetical signal distributions for $m_{Z'}=2\,\tev$ and $m_{Z'}=3\,\tev$ with arbitrary normalizations are plotted for illustration purposes.
The bottom panel gives the significance (deviation / statistical uncertainty) for each bin, calculated using the recommendation
of Ref. \cite{Choudalakis_2012}.
The impact on the background fit from the statistical uncertainties of BPDF parameters is shown as a light band around the solid line.
This effect is incorporated into the significance calculation.
(c) Observed and expected 95\% confidence-level limits on $\sigma\times \mathcal{B}$  as a function of $m_{Z'}$. }
\label{fig:result}
\end{center}
\end{figure*}
}
The signal from a hypothetical $Z'$ resonance is extracted as $\sigma\times \mathcal{B}$, defined as its production cross-section times the
decay branching fraction $\mathcal{B}(Z'\to H\gamma)$, by performing an unbinned
extended maximum-likelihood fit to the $m_{J\gamma}$ distributions of the selected events in the double- and single-$b$-tagged categories.
The predicted SM value of the $H\to b\bar{b}$ decay branching ratio, $0.582\pm 0.007$~\cite{deFlorian:2016spz}, is
used to calculate the upper limit on $\sigma\times \mathcal{B}$ from $\sigma\times \mathcal{B}(Z'\to H\gamma)\times \mathcal{B}(H \to b\bar{b})$. 
The fitting range for the double-$b$-tagged category is from $0.6\,\tev$ to $4.2\,\tev$, while for the single-$b$-tagged category
it is from $1.4\,\tev$ to $4.2\,\tev$ because of poor sensitivity in the low mass region.
Systematic uncertainties are taken into account as nuisance parameters with Gaussian sampling distributions~\cite{EXOT-2016-30}.
The lowest local (global) $p$ value is 0.005 (0.412) at $775\,\gev$, corresponding to a local (global) significance of  2.6$\sigma$ (0.22$\sigma$).
No significant signal-like excess is observed and the data are found to be described very well by a background-only fit,
as shown in Figs.~\ref{fig:result}(a) and \ref{fig:result}(b).
Hypothetical signal distributions for $m_{Z'}=2\,\tev$ and $m_{Z'}=3\,\tev$ with arbitrary normalizations are also plotted in Figs~\ref{fig:result}(a) and \ref{fig:result}(b) for illustration purposes.
Combined upper limits on the signal $\sigma\times \mathcal{B}$ at the 95\% confidence level are derived using  a modified frequentist 
method~\cite{Junk:1999kv,Read:2002hq}, with toy MC experiment, taking into account both the statistical and systematic uncertainties. The result as a function of the resonance mass
is shown in Fig~\ref{fig:result}(c).
The better sensitivity and larger integrated luminosity ($139\,\ifb$) of this search lowers the expected upper limits of this search
as compared to that of the previous ATLAS search ($36.1\,\ifb$)~\cite{EXOT-2016-30}.
The ratio of the current expected upper limits to that of the previous result is about 1/3\,(1/15) for resonances with masses below $1.2\,\tev$\,(above 2.5\,\tev).
A similar comparison with that of the previous CMS search ($35.9\,\ifb$)~\cite{CMS-EXO-17-019},
where a multivariable approach based on a boosted decision tree was used to identify $H\to b\bar{b}$ decays,
finds a ratio that varies between 2/5 and 1/3 for masses below 2.5\,\tev.
 
In conclusion, this Letter reports on a search for the production of new heavy resonances decaying into a Higgs boson and a photon, using
$139~\mbox{fb}^{-1}$ of $\sqrt{s}=13\,\tev$  $pp$ collision data collected by the ATLAS detector at the LHC.
The analysis is performed by reconstructing the
hadronic decay of the Higgs boson as a single large-radius jet, targeting the $H\to b\bar{b}$ mode.
A novel algorithm using information about the jet constituents in the center-of-mass frame of the jet is implemented to identify the two
$b$ quarks in the jet and enhances the sensitivity of the search.
No significant excess of events is observed above the expected background. Upper limits are set on the production cross-section times branching fraction for
resonance decays into a Higgs boson and a photon in the resonance mass range from $0.7$ to $4\,\tev$, which is significantly wider than in
the previous ATLAS and CMS searches.


We thank CERN for the very successful operation of the LHC, as well as the
support staff from our institutions without whom ATLAS could not be
operated efficiently.
 
We acknowledge the support of ANPCyT, Argentina; YerPhI, Armenia; ARC, Australia; BMWFW and FWF, Austria; ANAS, Azerbaijan; SSTC, Belarus; CNPq and FAPESP, Brazil; NSERC, NRC and CFI, Canada; CERN; ANID, Chile; CAS, MOST and NSFC, China; COLCIENCIAS, Colombia; MSMT CR, MPO CR and VSC CR, Czech Republic; DNRF and DNSRC, Denmark; IN2P3-CNRS and CEA-DRF/IRFU, France; SRNSFG, Georgia; BMBF, HGF and MPG, Germany; GSRT, Greece; RGC and Hong Kong SAR, China; ISF and Benoziyo Center, Israel; INFN, Italy; MEXT and JSPS, Japan; CNRST, Morocco; NWO, Netherlands; RCN, Norway; MNiSW and NCN, Poland; FCT, Portugal; MNE/IFA, Romania; JINR; MES of Russia and NRC KI, Russian Federation; MESTD, Serbia; MSSR, Slovakia; ARRS and MIZ\v{S}, Slovenia; DST/NRF, South Africa; MICINN, Spain; SRC and Wallenberg Foundation, Sweden; SERI, SNSF and Cantons of Bern and Geneva, Switzerland; MOST, Taiwan; TAEK, Turkey; STFC, United Kingdom; DOE and NSF, United States of America. In addition, individual groups and members have received support from BCKDF, CANARIE, Compute Canada, CRC and IVADO, Canada; Beijing Municipal Science \& Technology Commission, China; COST, ERC, ERDF, Horizon 2020 and Marie Sk{\l}odowska-Curie Actions, European Union; Investissements d'Avenir Labex, Investissements d'Avenir Idex and ANR, France; DFG and AvH Foundation, Germany; Herakleitos, Thales and Aristeia programmes co-financed by EU-ESF and the Greek NSRF, Greece; BSF-NSF and GIF, Israel; La Caixa Banking Foundation, CERCA Programme Generalitat de Catalunya and PROMETEO and GenT Programmes Generalitat Valenciana, Spain; G\"{o}ran Gustafssons Stiftelse, Sweden; The Royal Society and Leverhulme Trust, United Kingdom.
 
The crucial computing support from all WLCG partners is acknowledged gratefully, in particular from CERN, the ATLAS Tier-1 facilities at TRIUMF (Canada), NDGF (Denmark, Norway, Sweden), CC-IN2P3 (France), KIT/GridKA (Germany), INFN-CNAF (Italy), NL-T1 (Netherlands), PIC (Spain), ASGC (Taiwan), RAL (UK) and BNL (USA), the Tier-2 facilities worldwide and large non-WLCG resource providers. Major contributors of computing resources are listed in Ref.~\cite{ATL-SOFT-PUB-2020-001}.
 

\printbibliography
\clearpage \input{atlas_authlist}

\end{document}

%% file: atlas_authlist.tex
 
\begin{flushleft}
\hypersetup{urlcolor=black}
{\Large The ATLAS Collaboration}

\bigskip

\AtlasOrcid[0000-0002-6665-4934]{G.~Aad}$^\textrm{\scriptsize 102}$,    
\AtlasOrcid[0000-0002-5888-2734]{B.~Abbott}$^\textrm{\scriptsize 128}$,    
\AtlasOrcid{D.C.~Abbott}$^\textrm{\scriptsize 103}$,    
\AtlasOrcid[0000-0002-2788-3822]{A.~Abed~Abud}$^\textrm{\scriptsize 36}$,    
\AtlasOrcid[0000-0002-1002-1652]{K.~Abeling}$^\textrm{\scriptsize 53}$,    
\AtlasOrcid[0000-0002-2987-4006]{D.K.~Abhayasinghe}$^\textrm{\scriptsize 94}$,    
\AtlasOrcid[0000-0002-8496-9294]{S.H.~Abidi}$^\textrm{\scriptsize 167}$,    
\AtlasOrcid[0000-0002-8279-9324]{O.S.~AbouZeid}$^\textrm{\scriptsize 40}$,    
\AtlasOrcid{N.L.~Abraham}$^\textrm{\scriptsize 156}$,    
\AtlasOrcid[0000-0001-5329-6640]{H.~Abramowicz}$^\textrm{\scriptsize 161}$
% ...
\bigskip
\\

% ...
$^{36}$CERN, Geneva; Switzerland.\\
% ...
$^{40}$Niels Bohr Institute, University of Copenhagen, Copenhagen; Denmark.\\
% ...
$^{53}$II. Physikalisches Institut, Georg-August-Universit\"{a}t G\"ottingen, G\"ottingen; Germany.\\
% ...
$^{94}$Department of Physics, Royal Holloway University of London, Egham; United Kingdom.\\
% ...
$^{102}$CPPM, Aix-Marseille Universit\'e, CNRS/IN2P3, Marseille; France.\\
$^{103}$Department of Physics, University of Massachusetts, Amherst MA; United States of America.\\
% ...
$^{128}$Homer L. Dodge Department of Physics and Astronomy, University of Oklahoma, Norman OK; United States of America.\\
% ...
$^{156}$Department of Physics and Astronomy, University of Sussex, Brighton; United Kingdom.\\
% ...
$^{161}$Raymond and Beverly Sackler School of Physics and Astronomy, Tel Aviv University, Tel Aviv; Israel.\\
% ...
$^{167}$Department of Physics, University of Toronto, Toronto ON; Canada.\\
% ...

\end{flushleft}
